# A Carrington-like geomagnetic storm observed in the 21st century

C. Cid[1], E. Saiz[1], A. Guerrero[1], J. Palacios[1] and Y. Cerrato[1]

[1]Space Research Group - Space Weather, Departamento de Física y Matemáticas, Universidad de Alcalá, Alcalá de Henares, Spain (consuelo.cid@uah.es).

**Abstract.** In September 1859 the Colaba observatory measured the most extreme geomagnetic disturbance ever recorded at low latitudes related to solar activity: the Carrington storm. This paper describes a geomagnetic disturbance case with a profile extraordinarily similar to the disturbance of the Carrington event at Colaba: the event on 29 October 2003 at Tihany magnetic observatory in Hungary. The analysis of the *H*-field at different locations during the "Carrington-like" event leads to a re-interpretation of the 1859 event. The major conclusions of the paper are the following: (a) the global *Dst* or *SYM-H*, as indices based on averaging, missed the largest geomagnetic disturbance in the 29 October 2003 event and might have missed the 1859 disturbance, since the large spike in the horizontal component (*H*) of terrestrial magnetic field depends strongly on magnetic local time (MLT); (b) the main cause of the large drop in *H* recorded at Colaba during the Carrington storm was not the ring current but field-aligned currents (FACs), and (c) the very local signatures of the *H*-spike imply that a Carrington-like event can occur more often than expected.

## 1. Introduction

Major disturbances of the magnetosphere, or geomagnetic storms, are a consequence of solar activity, and represent a serious hazard to our technology dependent society. On 2 September 1859, the Colaba observatory measured the most extreme geomagnetic disturbance ever recorded at low and mid latitudes. Although at that time the magnetic observatories used to produce *Dst* were not in operation, the *Dst* minimum value for this event was first estimated by *Siscoe* [1979] about −2000 nT. More recently, *Lakhina et al.* [2005] and *Cid et al.* [2013] provided new estimations of −1760 nT and −685 nT, respectively. In any case, the so-called Carrington storm is ranked, according to the *Dst* index, as the most extreme geomagnetic storm ever recorded.

A large amount of guesswork that has been done concerning the Carrington storm: inferring the estimation of the minimum *Dst* index just from one observatory [*Akasofu and Kamide*, 2005; *Cid et al.*, 2013; *Li et al.*, 2006; *Siscoe et al.*, 2006; *Siscoe*, 1979; *Tsurutani et al.*, 2003, 2005]; estimating the flare intensity from its ground effect [*Cliver and Svalgaard*, 2004; *Clarke et al.*, 2010; *Boteler*, 2006]; identifying the interplanetary trigger (ICME, sheath, successive ejections) [*Tsurutani et al.*, 2003, 2005; *Siscoe et al.*, 2006; *Manchester et al.*, 2006]; discussing the relevance of ionospheric versus magnetospheric effects on the disturbance [*Akasofu and Kamide*, 2005; *Siscoe et al.*, 2006; *Green and Boardsen*, 2006; *Tsurutani et al.*, 2005], etc.

In October 2003, during the first Halloween storm, Tihany magnetic observatory (THY: MLat 45.87º, MLong 100.06º) in Hungary recorded a geomagnetic disturbance with an extraordinarily similar profile to that recorded at Colaba in 1859, but this time with a large dataset available from modern solar, interplanetary and terrestrial surface observatories. Although there is a large difference between the magnetic latitude of Colaba (~10º) and Tihany (~46º), both observatories were located within 10º of the lowest limit of the aurora observing locations for the corresponding event [*Green and Boardsen*, 2006; *Pallamraju and Chakrabarti*, 2005].

This paper is dedicated to the analysis of the geospace disturbances during the event on 29 October 2003 for the interval when it can be compared to the Carrington storm, i.e., the first hours after the sudden commencement (SC) observed in *Dst* and *SYM-H* indices. After the analysis of the event with the large data sets available, we extrapolate the results to the Carrington storm, taking into account the limited data available for this event. Section 2 describes the discovery of a Carrington-like event in local magnetic records. In Section 3 we analyse the geospace disturbances during the event on 29 October 2003. Finally, Section 4 is dedicated to discussion and Section 5 to Conclusions.

## 2. Discovering a Carrington-like profile in local magnetic records

After a careful visual scrutiny of measurements recorded by INTERMAGNET geomagnetic observatories at mid-latitudes during extreme storms, we found that the profile of the horizontal magnetic field component intensity measured on 29 October 2003 by some magnetic observatories in Europe was similar to that recorded at Colaba on 2 September 1859. Specifically Tihany observatory, in Hungary, recorded magnetic field variations which closely resemble the Carrington storm (Figure 1): the local timing of the large drop in *H* was similar (~ 9 - 10 MLT) -which is the onset of the terrestrial disturbance-, the short duration of the drop in *H* was comparable (< 1 hour), and the recovery to quiet time levels was extremely fast (< 1 hour) and also similar in shape. Nevertheless, the maximum disturbance is just half the intensity (almost 800 nT).

## 3. Geospace disturbances during the event on 29 October 2003

Figure 2 shows the terrestrial disturbance during the first hours of the storm event on 29 October 2003 (see Table 1 for geographical and magnetic coordinates of the observatories). Measurements of the horizontal magnetic field component recorded at mid-latitude (~40º MLat) observatories spread in longitude appear in Figure 2a. Panels from top to bottom are: SUA, IRT, FRN, BSL, and SPT (increasing in MLong). Figure 2b shows, for the same interval as Figure 2a, the *SYM-H* and *Dst* indices (top panel), and the horizontal magnetic field component recorded at the four low-latitude observatories involved in the computation of the *Dst* index: HER, KAK, HON, and SJG.

Regarding the *SYM-H* index, the first magnetic signature of the storm is a sudden commencement of approximately 80 nT (solid vertical line at 06:12 UT), which can be also observed in the *Dst* index. After the SC, *SYM-H* index decreases to a minimum value of -432 nT at 23:03 UT on 30 October (not shown in the figure) in a three-step profile. The SC signature is commonly related to an enhancement of the magnetopause current, which is expected to be observed across the globe. Therefore, we proceed to carefully analyse magnetic records at different locations.

First we compare the measurements of the horizontal component from a chain of mid-latitude magnetic observatories spread in longitude in order to provide a sample all around the globe. The magnetic trace at IRT, located in the mid-latitude afternoon sector, shows an increase of just less than 70 nT at SC onset. However, larger enhancements of ~120 nT appear at FRN and BSL at the SC. FRN and BSL, in the pre- and mid-night sector located, show a positive change which reaches ~300 nT and ~350 nT respectively, about one hour after SC. SPT and SUA, located in the dawn and pre-noon sector, show different behaviour: instead of a positive disturbance, a rapid decrease appears at the time of the SC leading to negative depression development similar to the Carrington event and to that shown in Figure 1 for THY, herein after *H*-spike.

The intensity of the disturbance decreases rapidly when moving in longitude from dawn (where THY is located at the peak of the *H*-spike) (~25 %, 5 degrees away from THY at SUA, and ~40 %, 20º further away at SPT). These observations reveal that the extreme drop in *H* recorded at some observatories on 29 October 2003 was actually a very local drop in the dawn sector.

At low latitudes, we have checked the magnetic traces for this event at the four observatories involved in computing the *Dst* index (HER, KAK, HON and SJG). The SC appears at HON, KAK and SJG but is not noticeable at HER (in the dawn sector). Then, the disturbance during the first two hours after the SC depends strongly on the sector at low latitudes as previously shown for mid-latitudes. The horizontal component shows a negative depression at HER reaching a minimum value of -330 nT, while the disturbance goes to positive values at SJG (in the after mid-night sector), reaching a maximum value of 210 nT at the time of the maximum disturbance at HER. KAK and HON show smaller disturbances, but also opposite (positive in HON and negative in KAK). The result of an average of all these measurements in the computation of the *Dst* index is the almost complete disappearance of the disturbance during these two hours after the SC. The higher temporal resolution of *SYM-H* index does not prevent from missing the disturbance of these first two hours after the SC, as this index is also computed by averaging disturbances at different longitudes.

From the third hour to the minimum value of the disturbance at 00:33 UT on 30 October, a slower decrease of the horizontal component is a common signature at all magnetic observatories analysed at low- and mid-latitude. Therefore, *Dst* and *SYM-H* indices are good indicators of the disturbance for this time.

Measurements by the DMSP satellite (F13) at 17:45 local time, reported by *Mannucci et al.* [2005], showed that the drift velocity at DMSP altitude (840 km) was predominantly negative (downward) during the days 25–30 October 2003, except for three distinct periods when the drift velocity becomes significantly upward (>70 m s$^{-1}$): 29 October at 06:16 UT and at 19:51–23:13 UT; and 30 October at 19:41–21:18 UT. Large FACs were observed by CHAMP in the dayside in both hemispheres between ~6-8 UT [*Wang et al.*, 2006].

Enhancements in estimated hemispheric power index [*Fuller-Rowell and Evans*, 1987; *Emery et al.*, 2008] from POES in Northern hemisphere were observed from ~6 UT on 29 October until ~5 UT on 30 October, reaching up to 554.5 GW on 29 October at 07:19 UT. The hemispheric power index was enhanced again on the late hours on 30 October. These data reveal enhancement and prolonged auroral activity which was already observed using OI 6300 Å emission from Boston (48.3º MLat), which is approximately at 30º-40º equatorward in latitude of the normal auroral precipitation regions [*Pallamraju and Chakrabarti*, 2005]. NOAA Technical Memorandum OAR SEC-88 on Halloween storms reported that aurora sightings occurred from California to Houston to Florida. Tremendous aurora viewing was also reported from mid-Europe and even as far south as the Mediterranean countries (~40º MLat).

## 4. Discussion

We report in this paper a geomagnetic disturbance case recorded on 29 October 2003 at THY magnetic observatory (hereafter, C03), whose temporal profile during two days is extraordinarily similar to the one recorded at Colaba in 1859 (hereafter, C59). Of course there are two remarkable differences among the Carrington event (C59) and the "Carrington-like" event (C03): (1) the intensity of the disturbance in C59 is double that in C03, and (2) the latitude of the magnetic observatory where the measurements were collected, which was very low for C59 (10º MLat) and middle for C03 (46º MLat). However, those latitudes have a common feature: both were about 10º equatorward from the lower latitude reported of aurora sightings for the corresponding event, which in both cases was far from the typical edge of the auroral oval.

Comparison between horizontal magnetic field component measurements at low-latitude observatories during C03 (Figure 2b) reveals that the size of the disturbance in the first two hours following the SC depends strongly on the local time of the magnetic records, being negative in the dayside and positive in the nightside. A small increase appeared in *Dst* or *SYM-H*, corresponding to the SC, and the largest disturbance (the *H*-spike) disappeared in these global indices due to the averaging involved in the computation procedure.

The problem for *Dst* or *SYM-H* which missed the largest disturbance is not related in this case to the latitude of the magnetic observatories involved in the computation of the indices, as it usually occurs in high latitude events. Computing global indices at the time of the *H*-spikes from an average of *H* from mid-latitude observatories widely spread in longitude also results in values close to quiet time for those indices due to the large asymmetry in *H* records (extreme positive and negative disturbances are averaged out). Therefore, global indices, commonly used for space weather purposes, based on the mean value of the magnetic disturbance levels, as *Dst* and *SYM-H*, cannot be considered as good proxies to quantify the disturbance, as both might miss one of the most hazardous ground effects. This result agrees with *Akasofu and Kamide* [2005] that believed that it is not likely that the C59 storm reached a *Dst* decrease of as large as 1760 nT. Our results also agree with *Li* [2009] and *Cid et al.* [2014] on the relevance of local magnetic field disturbances against a global activity level for technological systems. Moreover, considering the very local signatures of the *H*-spike and the lack of magnetic observatories covering all MLT sectors in past epochs, our outcomes also imply that a Carrington-like event can occur more often than expected, as suggested by *Kataoka* [2013].

As previously shown for low-latitude magnetic observatories, longitudinal analysis of magnetic field measurements at mid-latitude observatories during C03 (Figure 2a) shows negative or positive depressions depending on the local time. These depressions happened simultaneously with auroral effects. The longitudinal variation at mid-latitudes of the geomagnetic disturbance during C03 *H*-spike can be summarized as a day-night asymmetry peaking at dawn-noon sector and almost unnoticed in the noon-dusk sector, although the peak disturbance due to an asymmetric ring current effect is expected in the dusk sector [*Li*, 2009]. This result does not agree with *Tsurutani et al.* [2003] or *Li et al.* [2006] which pointed out that the *H*–spike at Colaba was related to the plasma injection into the nightside magnetosphere, which enhances the ring current. However, the spike happened from ~ 9 to 10 MLT, where other currents have a major contribution (*Shi et al.*, 2008). *Akasofu and Kamide* [2005] pointed out that some intense storms tend to have a sharp forenoon decrease of the *H* component, which accompanies a greater disturbance at auroral latitudes. These signatures were previously related to FACs by *Tsuji et al.* [2012] for the geomagnetic storm on 7-8 September 2002 and by *Yu et al.* [2010] using MHD simulations.

*Love and Gannon* [2010] also reported a large dawn-dusk asymmetry in low-latitude disturbances during the November 2003 storm and during the October 2003 storm. They proposed that these observations were the result of the superposition of magnetopause currents and partial ring current and, even FACs. Similar evidence of magnetic changes, confined in a narrow longitude range in the forenoon sector, and as a result with a little contribution to the *Dst* index, were provided by *Akasofu and Kamide* [2005] for some intense storms. *Siscoe et al.* [2006] and *Green and Boarden* [2006] already suggested that the *H*-spike in C59 might have a significant auroral-ionospheric component. *Boteler* [2006] pointed out that the combination of the negative depression in *H* at Rome and the large positive *Z* excursion at Greenwich were the signatures of a large westward electrojet at ~40º to 45º MLat, which will produce a considerable enlargement of the auroral oval during C59. However, he concluded that there was no evidence that it went far enough equatorward to contribute to the magnetic variation recorded at Colaba. More recently *Cliver and Dietrich* [2013], comparing C59 and the geomagnetic storm in May 1921, indicated that an auroral negative depression contributed to the negative *H*-spike in the Colaba trace during C59.

During C03, the availability of a large number of magnetic records spread through the globe has allowed us to go further in these suggestions regarding the auroral-ionosphere component in magnetograms and to propose a major role for FACs in C03 and by extrapolation during C59. Extending what *Siscoe et al.* [2006] pointed out for the effect of ionospheric currents involved in C59 to FACs, we conclude that the effect of FACs affecting a magnetogram at a latitude as low as 10º is also an exceptional aspect of the storm.

The *H*-spike at Colaba in C59 (considered up to now as a *Dst*-spike) was addressed by simulation assuming solar wind conditions with very intense (~ 60 nT) and short time duration southern $B_z$ and an additional extreme enhancement of the solar wind density (~1800 cm$^{-3}$) after the negative $B_z$ [*Li et al.*, 2006]. Similar profile shapes to the *Li et al.*

[2006] solar wind conditions have recently been seen in space plasma data [*Kozyra et al.*, 2013] and intense short-duration southern fields (~50 nT) were also recorded during 29 October 2003. However, extreme density values as those proposed by *Li et al.* [2006] have never been recorded before (as an example of extreme density event, we can cite the March 2001 ICME [*Farrugia et al.*, 2006] which had densities just after the shock in excess of 100 cm$^{-3}$). This extreme density value is a key-point in the simulation process as that is the only way to match the fast recovery of Colaba record assuming that the drop in *H* was due to the ring current. *Li et al.* [2006] modelled *Dst* as a sum of several terms, each one varying in time depending on a "source" term (which represents the external driving) and a "loss" term (which represents the decay rate of the field source). In the simulation, the term representing the contribution of the main ring current dominates the decrease of *Dst* and a pressure dependent term dominates the fast recovery of *Dst*. This simulation can be considered as a "dynamically driven" approach [see *Tsyganenko*, 2013 and references therein]. In every dynamically driven approach the loss term depends on which current system is being considered. The TS05 magnetosphere model [*Tsyganenko and Sitnov*, 2005] provides interesting information regarding the dynamics and peak values of the total current corresponding to individual sources. Their relaxation/response timescales were found to differ significantly from each other, from as large as ~30 hours for the symmetric ring current to only ~50 min for the FACs. In C03 the time between the minimum value of *H* and the end of the spike was 45 min and in C59 was ~50 min. Moreover, FACs peak at prenoon sector (~9 MLT), which was the location of Colaba for C59 and THY for C03. All these results point to FACs as the main current involved in the large *H*-spikes of the "Carrington-like" events.

**5. Conclusions**

The analysis of local magnetic records and magnetic indices during the event on 29 October 2003, when they are extrapolated to the Carrington event, led us to interesting implications. These are the following:

1) The *Dst* or *SYM-H* indices (commonly used to assess the severity of a storm) if calculated with multiple observatories as done today might have missed the large *H*-spike recorded at Colaba during C59.

2) The very local signatures of the *H*-spike imply that a Carrington-like event can occur more often than expected.

3) FACs played a major role in the *H*-spike in C03 and C59.

**Acknowledgements**. Geomagnetic field data have been obtained from INTERMAGNET magnetic observatories. The authors thank the Institutes that operate the observatories which provided data for this study; they also acknowledge the WDC for Geomagnetism for the magnetic field data and the *SYM-H* index.

This research was supported by the contract "Estudio de la influencia de fenómenos relacionados con la Meteorología Espacial en las infraestructuras de REE" between Red Eléctrica de España and the Universidad de Alcalá and by grants PPII10-0183-7802 from the Junta de Comunidades de Castilla-La Mancha of Spain and AYA2013-47735-P from MINECO. The editor thanks two anonymous referees for their assistance in evaluating this paper.

# References

Akasofu, S.-I., and Y. Kamide (2005), Comment on ''The extreme magnetic storm of 1–2 September 1859'' by B. T. Tsurutani, W. D. Gonzalez, G. S. Lakhina, and S. Alex, *J. Geophys. Res.*, 110, A09226, doi:10.1029/2005JA011005.

Boteler, D. H. (2006), The super storms of August/September 1859 and their effects on the telegraph system, *Adv. in Space Res.*, 38, 159-172.

Burton, R. K., R. L. McPherron, and C. T. Russell (1975), An empirical relationship

Cid, C., J. Palacios, E. Saiz, Y. Cerrato, J. Aguado, and A. Guerrero (2013), Modeling the recovery phase of extreme geomagnetic storms, *J. Geophys. Res.*, 118, doi:10.1002/jgra.50409.

Cid, C., J. Palacios, E. Saiz, A. Guerrero, Y. Cerrato (2014), On extreme geomagnetic storms, J. Space Weather Space Clim., 4, A28, doi: 10.1051/swsc/2014026.

Clarke, E., C. Rodger, M. Clilverd, T. Humphries, O. Baillie, and A. Thomson (2010), An estimation of the Carrington flare magnitude from solar flare effects (sfe) in the geomagnetic records, *Royal Astron. Soc. National Astron. Meeting*, 12–16 April, 2010, University of Glasgow, UK (available at http://nora.nerc.ac.uk/19904/).

Cliver, E. W., and W. F. Dietrich (2013), The 1859 space weather event revisited: limits of extreme activity, *J. Space Weather Space Clim.*, 3 A31, doi: 10.1051/swsc/2013053.

Cliver, E.W., and L. Svalgaard (2004), The 1859 solar-terrestrial disturbance and the current limits of extreme space weather activity, *Solar Phys.*, 224, 407–422.

Emery, B. A., V. Coumans, D. S. Evans, G. A. Germany, M. S. Greer, E. Holeman, K. Kadinsky-Cade, F. J. Rich, and W. Xu (2008), Seasonal, Kp, solar wind, and solar flux variations in long-term single-pass satellite estimates of electron and

ion auroral hemispheric power, *J. Geophys. Res.*, 113, A06311, doi:10.1029/2007JA012866.

Farrugia, C. J., V. K. Jordanova, M. F. Thomsen, G. Lu, S. W. H. Cowley, and K. W. Ogilvie (2006), A two-ejecta event associated with a two-step geomagnetic storm, *J. Geophys. Res.*, 111, A11104, doi:10.1029/2006JA011893.

Fuller-Rowell, T. J., and D. S. Evans (1987), Height-integrated Pedersen and Hall conductivity patterns inferred from the TIROS/NOAA satellite data, *J. Geophys. Res.*, 92, 7606–7618, doi:10.1029/JA092iA07p07606.

Green, J. L. and S. Boardsen (2006), Duration and extent of the great auroral storm of 1859, *Adv. in Space Res.*, 38, 130-135.

Kataoka, R. (2013), Probability of occurrence of extreme magnetic storms, *Space Weather*, 11, doi:10.1002/swe.20044, 214–218.

Kozyra, J. U., W. B. Manchester IV, C. P. Escoubet, S. T. Lepri, M. W. Liemohn, W. D. Gonzalez, M. W. Thomsen, and B. T. Tsurutani (2013), Earth's collision with a solar filament on 21 January 2005: Overview, *J. Geophys. Res. Space Physics,* 118, 5967–5978, doi:10.1002/jgra.50567.

Lakhina, G. S., S. Alex, B. T. Tsurutani, and W. D. Gonzalez. Research on Historical Records of Geomagnetic Storms. In K. Dere, J. Wang, and Y. Yan, eds., Coronal and Stellar Mass Ejections, vol. 226 of *IAU Symposium*, 3–15, 2005.

Li, Q., Y. Gao, J. Wang, and D.-S. Han (2009), Local differences in great magnetic storms observed at middle and low latitudes, *Earth Planets Space*, 61, 995-1001.

Li, X., M. Temerin, B. T. Tsurutani, and S. Alex (2006), Modeling of 1–2 September 1859 super magnetic storm, *Adv. Space Res.*, 38, doi:10.1016/j.asr.2005.06.070, 273-279.

Love, J. J., and J. L. Gannon (2010), Movie-maps of low-latitude magnetic storm disturbance, *Space Weather*, 8, S06001, doi:10.1029/2009SW000518.

Manchester, W. B., Ridley, A. J.; Gombosi, T. I.; Dezeeuw, D. L. (2006), Modeling the Sun-to-Earth propagation of a very fast CME, *Adv. Space Res.*, 38, 2, 253–262, doi:10.1016/j.asr.2005.09.044.

Mannucci, A. J., B. T. Tsurutani, B. A. Iijima, A. Komjathy, A. Saito, W. D. Gonzalez, F. L. Guarnieri, J. U. Kozyra, and R. Skoug (2005), Dayside global ionospheric response to the major interplanetary events of October 29–30, 2003 Halloween Storms, *Geophys. Res. Lett.*, 32, L12S02, doi:10.1029/2004GL021467.

Pallamraju, D., and S. Chakrabarti (2005), First ground-based measurements of OI 6300 Å daytime aurora over Boston in response to the 30 October 2003 geomagnetic storm, *Geophys. Res. Lett.*, 32, L03S10, doi:10.1029/2004GL021417.

Shi, Y., E. Zesta, and L. R. Lyons (2008), Modeling magnetospheric current response to solar wind dynamic pressure enhancements during magnetic storms: 1. Methodology and results of the 25 September 1998 peak main phase case, *J. Geophys. Res.*,113, A10218, doi:10.1029/2008JA013111.

Siscoe, G., N. Cooker, and C. R. Clauer (2006), *Dst* of the Carrington storm of 1859, *Adv. Space Res.*, 38, doi:10.1016/j.asr.2005.02.102, 173–179.

Siscoe, G.L. (1979), A quasi-self-consistent axially symmetric model for the growth of a ring current through earthward motion from a prestorm configuration, *Planet. Space Science* 27, 285–295.

Tsuji, Y., A. Shinbori, T. Kikuchi, and T. Nagatsuma (2012), Magnetic latitude and local time distributions of ionospheric currents during a geomagnetic storm, *J. Geophys. Res.*, 117, A07318, doi:10.1029/2012JA017566.

Tsurutani, B. T., and W. D. Gonzalez (1995), The future of geomagnetic storm predictions: Implications from recent solar and interplanetary observations, *J. Atmosph. Sol. Terr. Phys.*, 57, 1369-1384.

Tsurutani, B. T., W. D. Gonzalez, G. S. Lakhina, and S. Alex (2003), The extreme magnetic storm of 1–2 September 1859, *J. Geophys. Res.*, 108 (A7), 1268, doi:10.1029/2002JA009504.

Tsurutani, B. T., W. D. Gonzalez, G. S. Lakhina, and S. Alex (2005), Reply to comment by S.-I. Akasofu and Y. Kamide on ''The extreme magnetic storm of 1–2 September 1859,'' *J. Geophys. Res.*, 110, A09227, doi:10.1029/2005JA011121.

Tsyganenko, N. A., and M. I. Sitnov (2005), Modeling the dynamics of the inner magnetosphere during strong geomagnetic storms, *J. Geophys. Res.*, 110, A03208, doi:10.1029/2004JA010798.

Tsyganenko, N. A. (2013), Data-based modeling of the Earth's dynamic magnetosphere: a review, *Ann. Geophys.*, 31, 1745-1772, doi:10.5194/angeo-31-1745-2013.

Wang, H., Lühr, S. Y. Ma, J. Weygand, R. M. Skoug, and F. Yin (2006), Field-aligned currents observed by CHAMP during the intense 2003 geomagnetic storm events, *Ann. Geophys.*, 24, 311–324.

Yu, Y., A. J. Ridley, D. T. Welling, and G. Tóth (2010), Including gap region field-aligned currents and magnetospheric currents in the MHD calculation of ground-based magnetic field perturbations, *J. Geophys. Res.*, 115, A08207, doi:10.1029/2009JA014869.

**Table captions**

**Table 1.** Location of INTERMAGNET ground observatories used in this work. Geomagnetic coordinates are given by using the IGRF geomagnetic field model year 2000.

| Observatory | IAGA code | Geodetic latitude (deg) | Geodetic longitude (deg) | Magnetic latitude (deg) | Magnetic longitude (deg) |
|---|---|---|---|---|---|
| Narsarsuaq | NAQ | 61.16 | 314.56 | 69.90 | 38.41 |
| Fort Churchill | FCC | 58.76 | 265.91 | 67.93 | 327.91 |
| Abisko | ABK | 68.36 | 18.82 | 65.89 | 114.8 |
| Tihany | THY | 46.90 | 17.54 | 45.87 | 100.06 |
| Fresno | FRN | 37.09 | 240.28 | 43.42 | 304.93 |
| San Pablo Toledo | SPT | 39.55 | 355.65 | 42.69 | 75.88 |
| Surlari | SUA | 44.68 | 26.25 | 42.38 | 107.50 |
| Irkutsk | IRT | 52.27 | 104.45 | 41.68 | 176.8 |
| Stennis S. Center | BSL | 30.35 | 270.36 | 40.06 | 339.49 |
| San Juan | SJG | 18.11 | 293.85 | 28.2 | 6.02 |
| Kakioka | KAK | 36.23 | 140.18 | 26.99 | 208.50 |
| Honolulu | HON | 21.32 | 202.0 | 21.5 | 269.48 |
| Alibag | ABG | 18.62 | 72.87 | 9.91 | 145.96 |
| Hermanus | HER | -34.43 | 19.23 | -33.75 | 83.73 |

**Figure Captions and Figures**

**Figure 1.** Horizontal component of the magnetic field recorded at Colaba (Bombay) during the 2-3 September 1859 geomagnetic storm (left panel) and at Tihany (Hungary) during the 29-30 October 2003 geomagnetic storm (right panel).

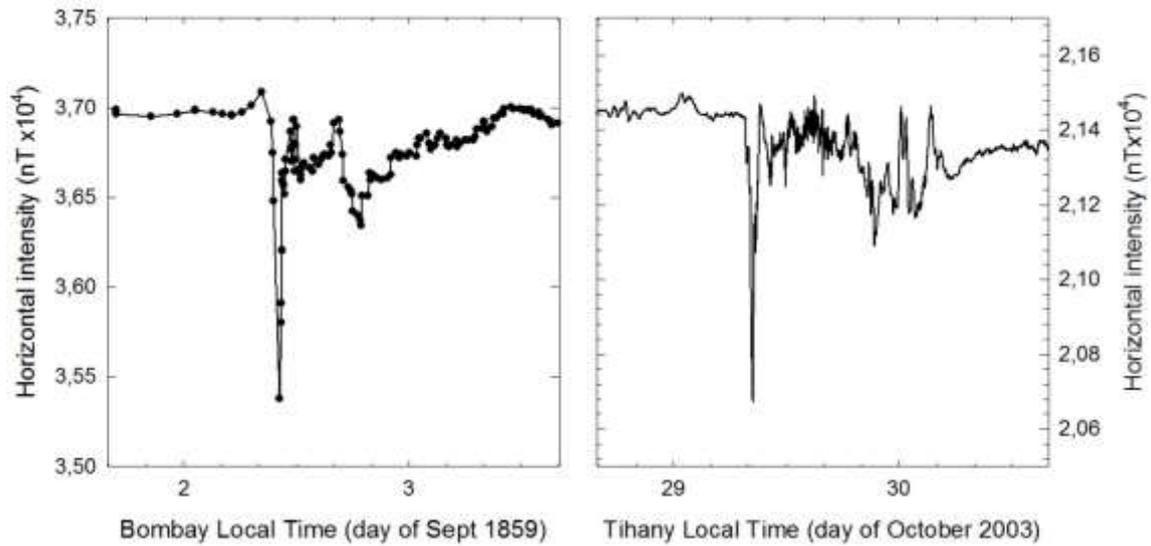

**Figure 2.** Geomagnetic disturbances of the 29 October 2003 event. Panel (a) shows measurements of the *H* component recorded at mid-latitude observatories. From top to bottom: SUA, IRT, FRN, BSL, and SPT (increasing in MLong). Panel (b), from top to bottom, displays *SYM-H* and *Dst* indices (top panel), and the *H* component recorded at the four low-latitude observatories involved in the computation of the *Dst* index: HER, KAK, HON, and SJG. Vertical solid line S marks shock discontinuity.

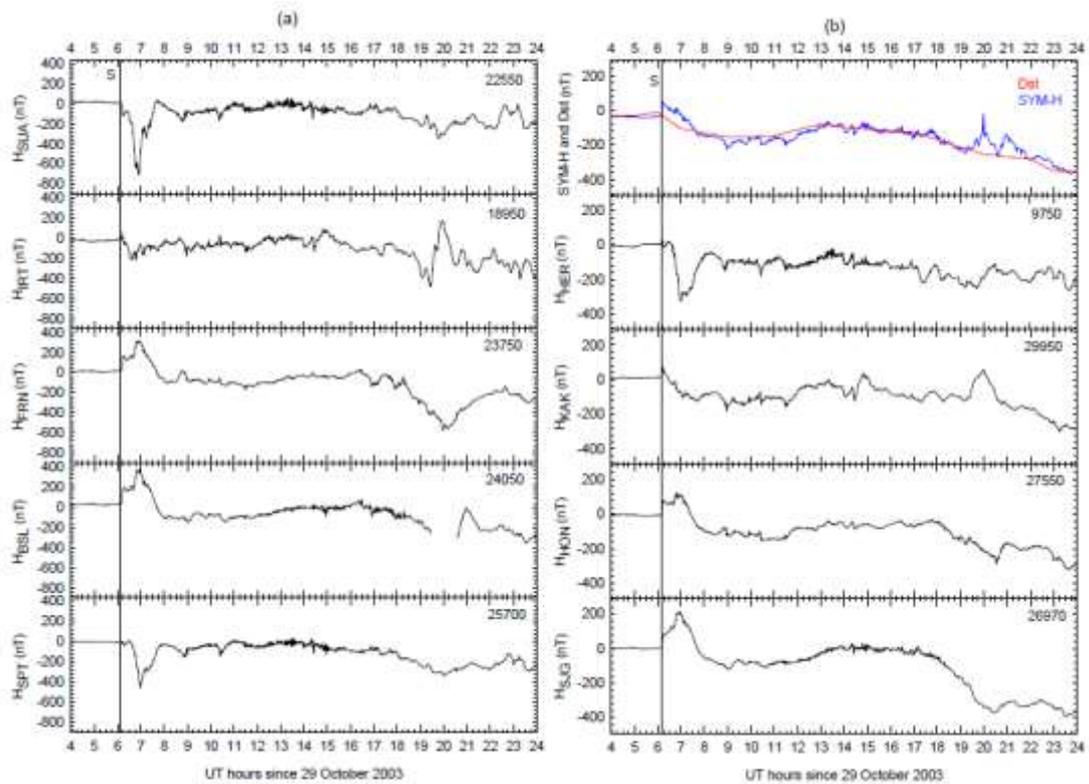